\documentclass[eqsecnum,aps,prb,twocolumn,showpacs,superscriptaddress]{revtex4}
\usepackage{graphicx}

\begin{document}
\title{Inhomogeneity effects in oxygen doped HgBa$_2$CuO$_{4}$}
\author{C. Ambrosch-Draxl} 
\email{cad@mu-leoben.at}
\affiliation{Institute for Physics, University Graz, Universit\"atsplatz 5, A-8010 Graz, Austria}
\affiliation{Chair of Atomistic Modelling and Design of Materials, University Leoben, Erzherzog-Johann-Stra\ss e 3, A-8700 Leoben, Austria}
\author{E. Ya. Sherman} 
\affiliation{Department of Physics, University of Toronto, 60 St.
George Street, Toronto, Ontario, Canada M5S 1A7}
\pacs{74.81.-g, 74.25.Jb, 74.72.-h}
\date{\today}

\begin{abstract}
We theoretically investigate inhomogeneity effects on the charges, electric field gradients and 
site-projected densities of states in HgBa$_2$CuO$_{4+\delta}$. We find pronounced differences 
in the doping-induced number of holes at different atomic sites. The contributions of these sites 
to the density of states in the vicinity of the Fermi level are peaked at the same energy, but 
vary in magnitude by up to 70 percent and have different energy dependence. 
Due to this energy dependence the role of the intrinsic inhomogeneities for superconductivity strongly 
depends on the energy and character of the quasiparticle mediating the Cooper pairing. 
Our results can explain the origin of doping-induced 
effects observed either by local or macroscopic experimental probes. 
\end{abstract}

\maketitle

\section{Introduction}

Doping is the most crucial parameter to influence the critical temperature in high $T_c$ cuprates. 
The metallic state is reached by doping through the replacement of ions or by introduction of excess oxygen. 
Thereby the dopant adds states at the Fermi level $E_F$, where these new charge carriers (holes) cause a 
downward shift of $E_F$ and the carrier concentration in the CuO$_2$ plane is changed. 

Most of the theoretical models make use of two simplifications to describe doping effects in the complex physics 
of these systems: First, they use the carrier concentration rather than the doping level as an input parameter, 
and second, they assume that the additional charge is uniformly distributed within the different sites of the same 
atomic species in the CuO$_2$ plane. Concerning the former point, it has been shown\cite{cad:03} for 
HgBa$_2$CuO$_{4+\delta} $ how doping influences the charge carrier distribution and what limits the amount of holes 
that can be created. For the latter point, no first-principles calculations have been performed for any of the cuprates.

At the same time there is strong experimental evidence for intrinsic inhomogeneous charge distribution in high 
$T_c$ compounds. For example, the existence of stripes \cite{Tranquada} in underdoped La$_2$CuO$_{4}$ and Bi based 
compounds is a clear manifestation of this. However, there are other sources of spatially non-uniform  charge density. 
Scanning tunneling microscopy (STM) experiments performed on Bi$_2$Sr$_2$CaCu$_2$O$_{8+x}$ exhibit an inhomogeneous 
surface carrier distribution on a length scale of 14 \AA, in the normal as well as in the superconducting state.\cite{Pan:01} 
It was claimed that this inhomogeneity has an unknown origin but is an intrinsic property of this doped material, i.e. not 
related to impurities. More recently, it was found that the nanoscale electronic disorder in cuprates can be traced
back to inhomogeneities on the atomic scale.\cite{McElroy:05}
Local bulk probes like nuclear quadrupole resonance (NQR) and nuclear magnetic resonance (NMR) experiments exhibit not only 
changes in the hole content with doping, but also the existence of different sites of the same species as it was seen in 
Tl based compounds \cite{Gerashenko:99} and Sr doped La$_2$CuO$_{4}$.\cite{Singer}

Since the spatial scale of inhomogeneities is determined by the amount of dopants, for doping levels in the regime of 
underdoped to optimally doped, investigations on a scale of a few lattice parameters are required.  Wang {\it et al.} 
considered the role of the Coulomb impurities located close to the CuO$_2$ plane on the spatial dependence 
of the gap in a $d$-wave superconductor within the $t-J$ model.\cite{Wang02}  In this work, we perform parameter-free 
calculations which allow a detailed analysis of the charge distribution. By this way the inhomogeneities can be quantitatively 
studied and linked to experimental findings. We investigate the doping-induced charge redistribution in HgBa$_2$CuO$_{4+\delta}$, 
which exhibits the highest $T_c$ among all single-layer cuprates. Focusing on the CuO$_2$ planes, we will show that the 
inhomogeneity is quite pronounced, it occurs on a scale of a few lattice constants, and is strongly doping dependent. 
We will also consider the consequences for the site-projected densities of states and their impact on 
superconductivity.  To this extent, we have performed a series of supercell calculations corresponding to oxygen
concentrations of $\delta$=1/8, 1/6, 2/9, and 1/4.

\section{Supercell calculations}

\begin{figure}[h]
\begin{center}
\includegraphics[width=3.7cm]{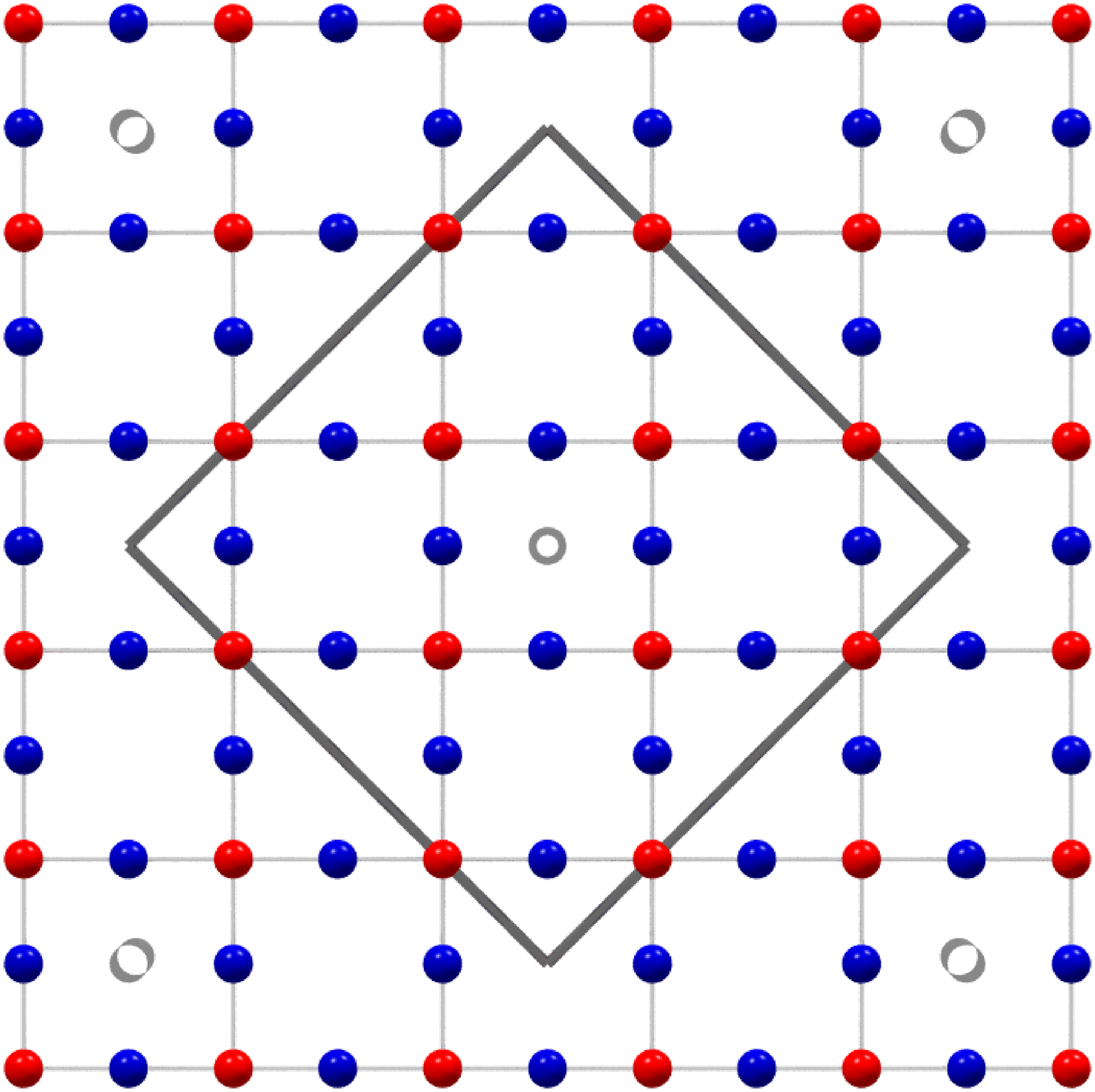}\qquad%
\includegraphics[width=3.7cm]{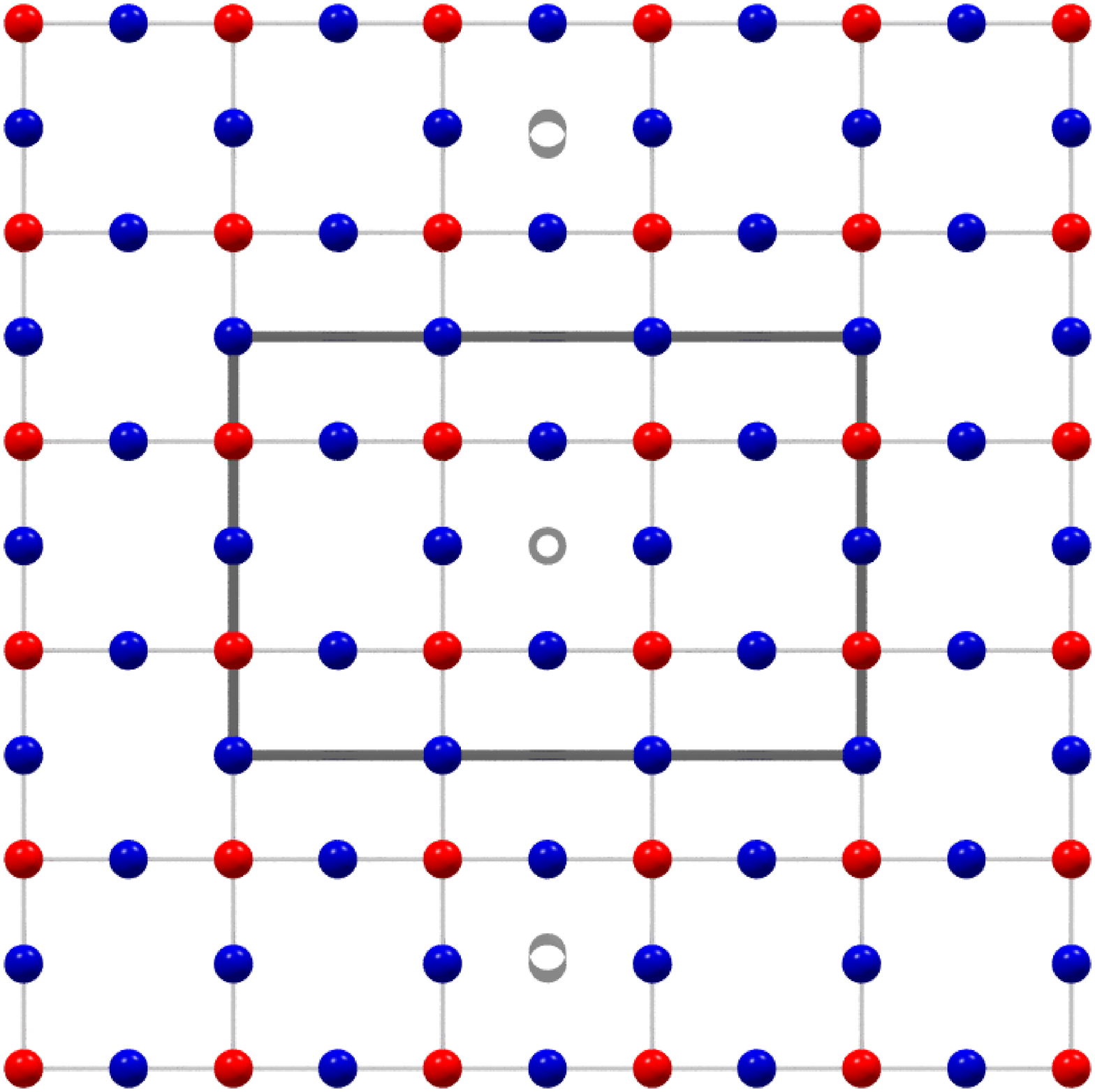}\\[0.5cm]
\includegraphics[width=3.7cm]{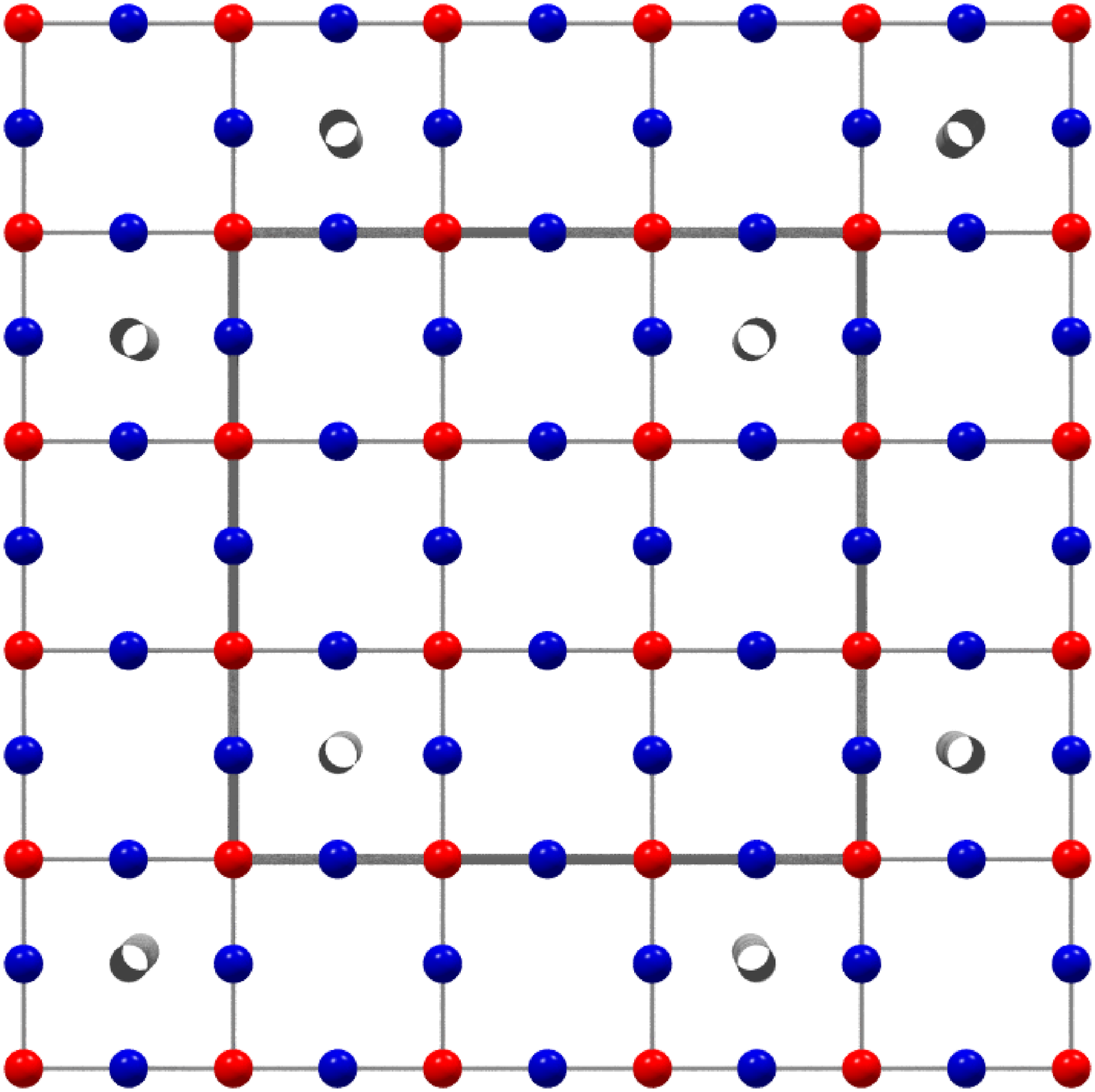}\qquad%
\includegraphics[width=3.7cm]{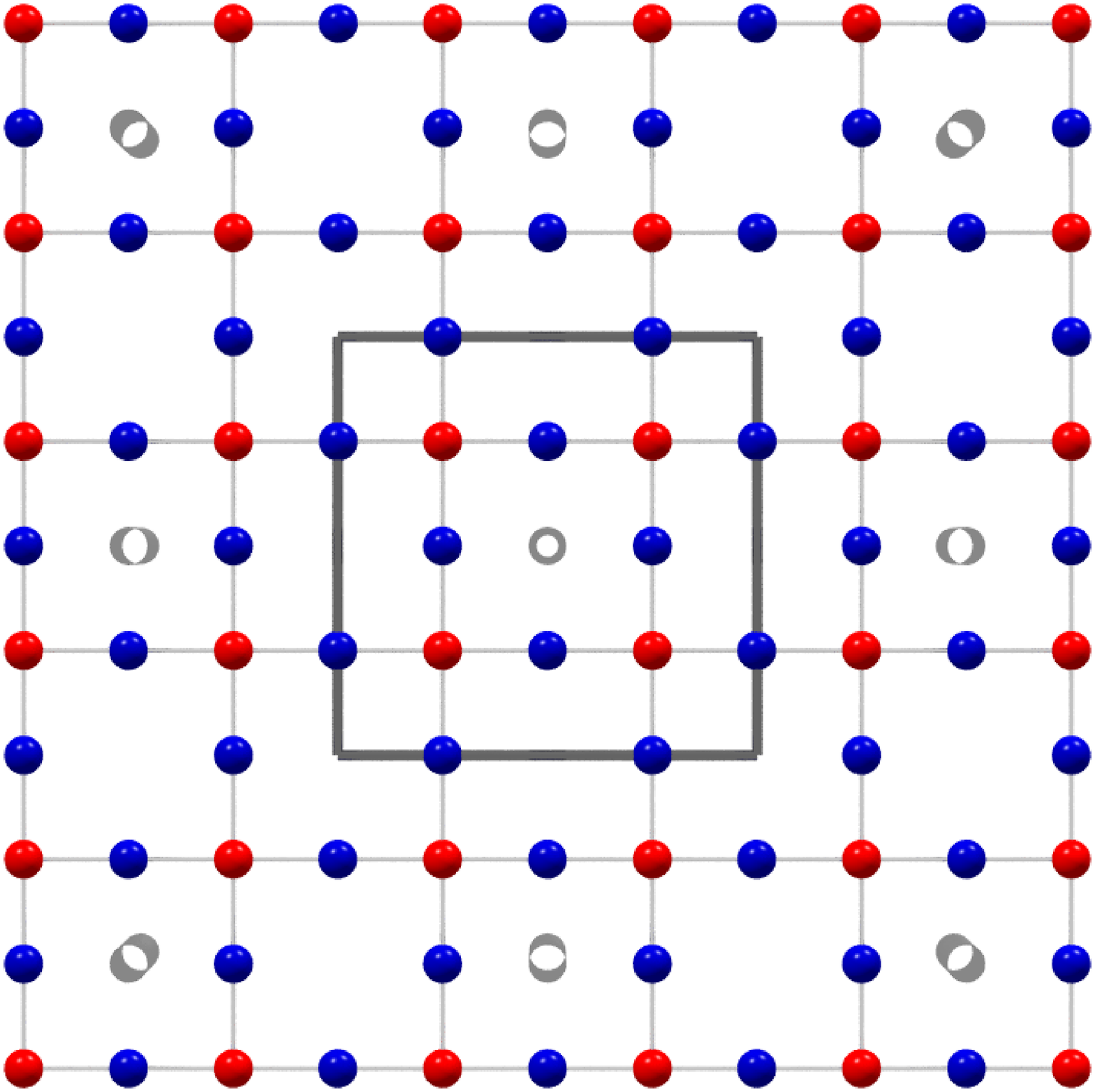}\\[0.5cm]
\hspace{4.2cm}
\includegraphics[width=3.7cm]{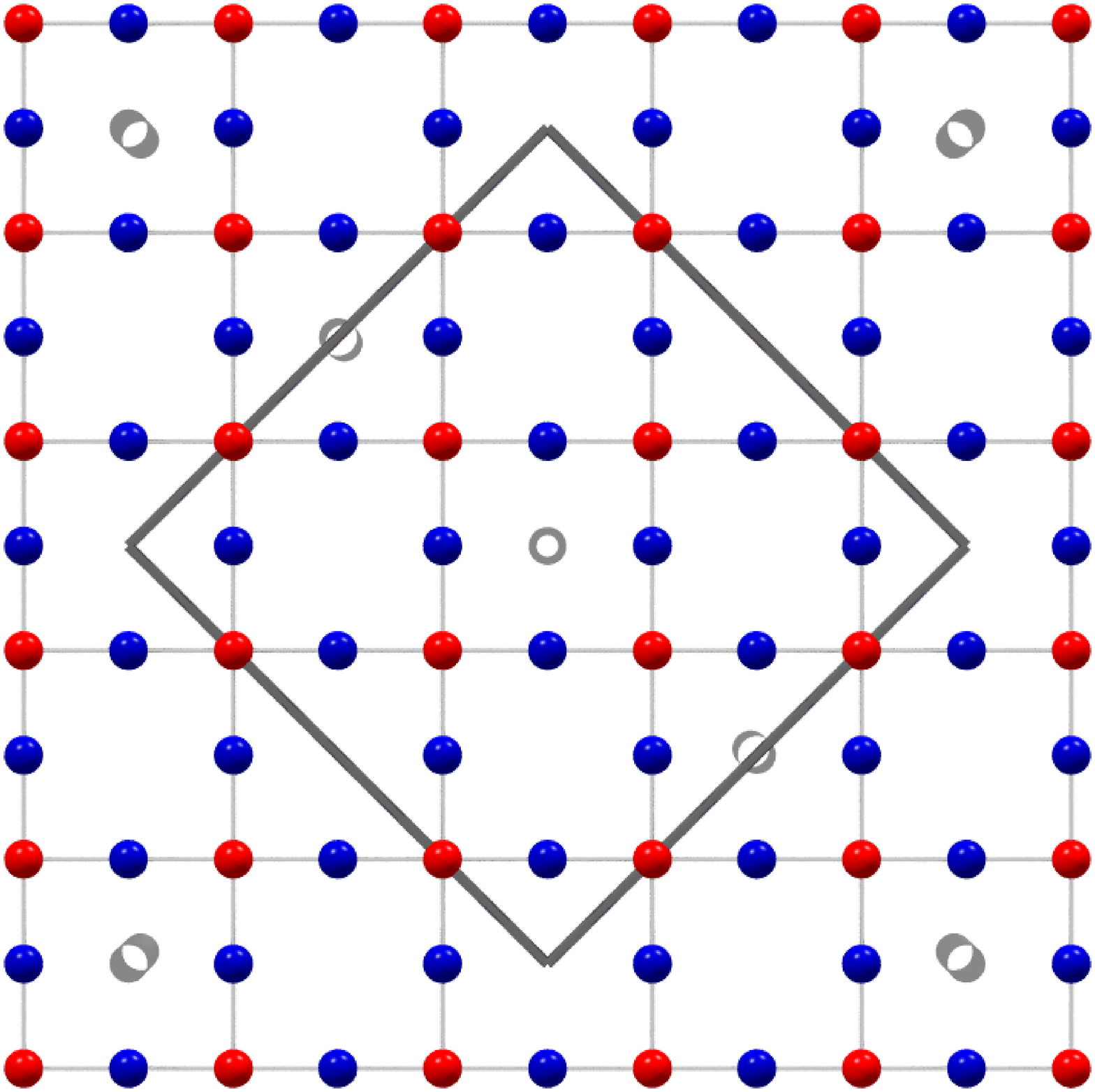}
\end{center}
\caption[]{(Color online) Copper-oxygen planes of the HgBa$_2$CuO$_{4+\delta}$ supercells representing the doping 
concentrations of $\delta$=0.125 (upper left right panel), 0.167 (upper right panel), 0.22 (lower left panel), 
and 0.25 (lower right panel).  The unit cells are indicated by the bold gray lines, the copper (oxygen) atoms by the 
red (blue) spheres. The projection of the dopant oxygen (located at the Hg plane) onto these planes is marked by the 
gray circles. In the lowest panel, an alternative structure for $\delta$=0.25 is displayed.}
\label{fig_supercells}
\end{figure}

Unit cells of the 8-, 6-, 9-, and 4-fold size compared to the undoped case are presented in Fig. \ref{fig_supercells}. 
The particular environment with respect to the dopant makes the various copper and oxygen sites within the CuO$_2$ planes 
inequivalent. The coordinates of these different types of Cu and O atoms are provided in Table \ref{t_holes} for the
structures shown in the two upper rows of the figure. The $\delta=0.167$ case gives rise to an orthorhombic cell with 
$3 \times 2$ single unit cell volumes, the three other structures are tetragonal.
 
The charge of a given site calculated within the supercell approach results from the influence of all dopant ions which 
may be generally long-ranged. However, due to the two-dimensional character of the carriers in the high-$T_c$ cuprates, 
the interaction of the dopant with its surrounding is screened at an in-plane length scale which is approximately the 
distance between the dopant and the CuO$_2$ plane.\cite{Ando82} Hence this interaction is short-ranged. Therefore the 
sizes of the chosen supercells are big enough to study the effects of charge redistribution caused even by a single 
dopant ion. The fact that the doping-induced charges averaged over the respective unit cell are only very weakly 
sensitive to the choice of the supercell has already been pointed out in Ref. \onlinecite{cad:03}. In order to get
more insight into the effects on the local environment, we consider an alternative supercell for $\delta=0.25$ as 
given in the lowest panel of Fig. \ref{fig_supercells}. This structure exhibits a less homogeneous distribution of dopants. 
Since it, however, is energetically unfavorable, as will be discussed below, the corresponding structural data and results are 
not displayed in Table \ref{t_holes}.   

All calculations have been carried out within density functional theory employing the full-potential linearized augmented 
plane-wave (LAPW) method utilizing the WIEN2k code.\cite{WIEN2k} Exchange and correlation effects are treated within the local 
density approximation (LDA). The computational details are given elsewhere.\cite{cad:03}

\section{Results}
\subsection{Charge distribution}

The atomic-like basis functions used within the LAPW method inside the so-called atomic spheres \cite{cad:lapw} allow to analyze the charge density within each sphere by decomposing it with respect to the contributions of different orbital characters. Although these numbers depend on the choice of the atomic-sphere radii they provide useful information about the charges around different atomic sites. 

\begin{table}
\renewcommand{\arraystretch}{1.2}
\begin{tabular}{| c | c | c  c | c  | c| c c c | c |}
\hline
doping & ion & X & Y & $d_O$ & $h_{s}$ [$e$] & $V_{11}$ & $V_{22}$ & $V_{33}$ & $\eta$\\
\hline
0.125  
& Cu1 (4)  & $\frac{1}{4}$ & 0               & $a\sqrt{10}$/2 
& 0.0515 &  2.6 &   2.6 & -5.2 & 0.01\\

& Cu2 (4)  & $\frac{1}{4}$ & $\frac{1}{2}$ & $a\sqrt2$/2 
& 0.0636 &  3.2 &  3.2 & -6.4 & 0.00\\

& O1 (4)  & $\frac{1}{8}$ & $\frac{1}{8}$ & 3$a$/2 & 0.0038 
& 14.2 & -8.8 & -5.4 & 0.24\\

& O2 (8)  & $\frac{1}{8}$ & $\frac{3}{8}$ & $a\sqrt5$/2 & 0.0089 
& 14.5 & -8.8 & -5.7 & 0.22\\

& O3 (4)    & $\frac{3}{8}$ & $\frac{3}{8}$ & $a$/2 & 0.0139
& 15.0 & -8.9 & -6.1 & 0.19\\
\hline
\hline
0.167 
& Cu1 (2)  & 0  & $\frac{1}{4}$              & $a\sqrt{10}$/2 & 0.0591
&  3.0 &   2.8 &  -5.8 & 0.02\\

& Cu2 (4)  & $\frac{1}{3}$ & $\frac{1}{4}$ &$a\sqrt{2}$/2 & 0.0730
&  3.5 &   3.5 &  -7.0 & 0.01\\    

& O1  (4)  & $\frac{1}{6}$ & $\frac{1}{4}$ & $a\sqrt5$/2 & 0.0087
& 14.7 &  -9.0 &  -5.7 & 0.22\\

& O2  (2)  & $\frac{1}{2}$ & $\frac{1}{4}$ & $a$/2        & 0.0195
& 15.5 &  -9.1 &  -6.4 & 0.18\\

& O3  (1)  & 0                  & $\frac{1}{2}$ & 3$a$/2      &0.0079
& 14.6 &  -9.0 &  -5.6 & 0.23\\
 
& O4  (2)  & $\frac{1}{3}$ & $\frac{1}{2}$ & $a$/2        & 0.0165
& 15.3 &  -9.1 &  -6.2 & 0.19\\
 
& O5  (1)  & 0                 & 0         & $a\sqrt{13}$/2 &0.0084
& 14.7 &  -9.0 &  -5.7 & 0.23\\
 
& O6  (2)  & $\frac{1}{3}$ & 0        & $a\sqrt5$/2     & 0.0121
& 14.9 &  -8.9 &  -6.0 & 0.20\\

\hline
\hline
0.22    
& Cu1 (1)  & 0                 & 0 & $a\sqrt2$/2 & 0.0863
&  3.6 &  3.9 & -7.5 & 0.04\\

& Cu2 (4)  & $\frac{1}{3}$ & 0 & $a\sqrt2$/2 & 0.0853
&  4.0 &  3.8 & -7.8 & 0.02\\

& Cu3 (2)  & $\frac{1}{3}$ & $\frac{1}{3}$ &$a\sqrt2$/2 & 0.0775
&  3.6 &  3.6 & -7.2 & 0.01\\

& Cu4 (2)  & $\frac{2}{3}$ & $\frac{1}{3}$ &$a\sqrt{10}$/2 & 0.0757
&  3.4 &  3.4 & -6.8 & 0.00\\

& O1  (4)   & $\frac{1}{6}$ & 0  & $a$/2 & 0.0253
& 15.7 & -9.2 & -6.6 & 0.16\\

& O2  (2)   & $\frac{1}{2}$ & 0  & $a\sqrt5$/2 & 0.0166
& 15.0 & -9.0 & -6.0 & 0.19 \\

& O3  (4)  & $\frac{1}{3}$ & $\frac{1}{6}$ & a/2 & 0.0241
& 15.6 & -9.2 & -6.4 & 0.17\\

& O4  (4)   & $\frac{2}{3}$ & $\frac{1}{6}$ &$a\sqrt{5}$/2 & 0.0164
& 15.1 & -9.1 & -6.0 & 0.21\\

& O5  (4)   & $\frac{1}{2}$ & $\frac{1}{3}$ & $a\sqrt{5}$/2 & 0.0151
& 15.0 & -9.1 & -5.9 & 0.21\\
\hline
\hline
0.25    
& Cu1 (4)   & $\frac{1}{4}$ & $\frac{1}{4}$ & $a\sqrt2$/2 & 0.0858
&  4.0 &  3.9 & -7.9 & 0.01\\

& O1  (4)    & $\frac{1}{4}$  & 0               & $a\sqrt5$/2 & 0.0190
& 15.4 & -9.0 & -6.4 & 0.17\\

& O2  (4)    & $\frac{1}{2}$  & $\frac{1}{4}$ & $a$/2          & 0.0218
& 15.7 & -9.1 & -6.6 & 0.16\\
\hline
\end{tabular}
\caption{Copper and oxygen types, their multiplicities, their (X,Y) coordinates given in lattice constants $A$ and $B$ of the respective supercells, and their in-plane distances to the nearest dopant oxygen $d_O$, where $a$ is the single cell lattice parameter. $h_{s}$ denotes the number of holes created by doping at the given site, $V_{\gamma\gamma}$ are the diagonal components of the electric field gradient tensor in 10$^{21}$ V/m$^2$ units, and $\eta$ is the corresponding asymmetry parameter. The principal axes of the EFG are either identical or close to the crystal axes.}
\label{t_holes}
\end{table}

The decrease in the partial charges in the atomic spheres averaged over the respective unit cells have been described in detail in Ref. \onlinecite{cad:03}. It was observed that a linear increase in the total number of holes for $\delta \le 0.22$, is followed by a plateau at higher $\delta$.  The reason for this saturation was found in the fact that at  $\delta\approx 0.22$ the dopant reaches a closed shell O$^{2-}$ configuration. The ratio of the additional average hole contents on Cu($d_{x^2-y^2}$) and O($p_x$) orbitals $h_{\rm Cu}/ h_{\rm O}$ is approximately $4.0$. (Note that there are two oxygen atoms in the CuO$_2$ plane per chemical formula unit, therefore the amount of oxygen holes is half that of copper.)

Now we focus on the charge distribution on the atomic scale. Table \ref{t_holes} presents the number of holes $h_{s}$ 
created at the specific copper and oxygen sites in the CuO$_2$ planes with the undoped material taken as the reference. 
As one can see, $h_{s}$ strongly depends on the atomic species, the specific site, and the doping content. Concerning the doping 
dependence, for all given $\delta$ values lower than $0.25$, the maximal difference in the number of doping-induced holes 
among non-equivalent sites of the same species (denoted as $\Delta h_{s}^{\rm max}$) is approximately 0.01 $e$, although the 
distances from the ions to the next dopant may be quite different at smaller doping concentrations. For a fixed doping content, 
we define a measure for the inhomogeneity, $\Delta h_{s}^{\rm max}$  divided by the average doping-induced charge of this species,
i.e. $h_{\rm Cu}$ or $h_{\rm O}$. This value decreases for copper from 21\% for $\delta$=0.125 to 13\% for $\delta$=0.22 (and 
zero for $\delta$=0.25). For oxygen it is much more pronounced with values of 114\%, 95\%, 51\%, and 14\% for $\delta$=0.125, 
0.167, 0.22 and 0.25, respectively. Summarizing these findings we conclude that the relative inhomogeneity with respect to the 
total amount of doping-induced holes (i) is more pronounced for oxygen than for copper, since doping introduces less holes on 
the oxygen than on the copper sites, and (ii) is decreasing with increasing $\delta$. 

The isotropy of the charge distribution for $\delta=0.25$ given in the table may be considered as an artifact due to the choice
of the supercell. For this reason, we have performed calculations for an alternative structure, which is displayed
in the lowest panel of Fig. \ref{fig_supercells}. For this case, the inhomogeneity is more pronounced, with 
$\Delta h_{s}^{\rm max}=0.013$ for copper, still being in the range of a hundredth of an electron as found for the other doping
concentrations. But more important, the total energy of this structure is more than 5 mRy per formula unit higher, 
which corresponds to a temperature of about 900K. This large energy compared to the other cell choice can be explained 
by the smaller distance between the nearest neighbor dopants, which, being highly negatively charged, repel each other. 
For the other doping concentrations, we have chosen the supercells such that the separation of dopants is maximal.  

The non-uniform charge distribution can also be understood in terms of the distances to the next dopant ion. Focusing on the 
projection onto the CuO$_2$ plane the spacings to the copper positions are $a\sqrt{2}$/2, $a\sqrt{10}$/2, etc. Although for all 
$\delta$ values the maximum amount of holes at Cu is created in the first coordination shell with the distance being 
$a\sqrt{2}$/2, its  value strongly depends on $\delta$. While for $\delta$ = 0.125 and 0.167 there are only two non-equivalent 
copper positions differing in the distance to the dopant, there are four non-equivalent Cu atoms for $\delta$ = 0.22. In this 
case, the biggest amount of holes is created at Cu1, which is influenced by two dopant sites with the same distance; Cu2 
exhibits the same spacing to the nearest dopant atom, but the next nearest dopant oxygen is farther away. A similar analysis 
can be done for the oxygen sites, where the in-plane distances to the dopant are  $a/2$,  $a\sqrt{5}$/2, 3$a$/2, etc. For 
example, at $\delta$ = 0.125, O3 (at a distance of $a$/2) is most affected by the excess oxygen, followed by O2, which interacts 
with a dopant at a distance of  $a\sqrt{5}$/2, and O1 where the dopant is 3$a/2$ away. Also for the oxygen sites the amount of 
created holes very strongly depends on the oxygen concentration, when positions with the same spacing to the dopant are compared. 
The values for the nearest dopant range from 0.0139 for $\delta$ = 0.125 via 0.0195 for $\delta$ = 0.167 to 0.0241 
for $\delta$ = 0.22.

\subsection{Electric field gradients}

The distribution of doping-induced charges can be probed by nuclear quadrupole and magnetic resonance. The electronic and nuclear 
charge densities form a non-uniform electric field within the unit cell. A nuclear quadrupole moment is coupled to the gradients 
of the electric field \cite{Cohen57} given by $V_{\alpha\beta}=\partial E_{\alpha}/\partial x_{\beta}$, where 
$E_{\alpha}$ is the component of the local electric field, and $x_{\beta}$  is the Cartesian coordinate. The traceless tensor
$V_{\alpha\beta}$ can be diagonalized and written in the principal axes as $V_{\gamma\gamma}$. The isotopes $^{17}$O and 
$^{63}$Cu are  usually used as the probes of electric field gradients in high-$T_c$ cuprates. As a result of the interaction 
of their nuclear quadrupole momenta $Q$ with the electric field gradients at the lattice sites, a resonance line at 
\begin{equation}  
\omega=\frac{eQ}{2\hbar}V_{zz}(1+{\eta^2}/3)^{1/2}
\label{quadrupole}
\end{equation}
appears in the radio frequency wave absorption spectrum of the system. Ordering the tensor elements by their absolute value, 
such that $V_{xx}$ is the smallest and $V_{zz}$ is the maximal component, the asymmetry parameter $\eta$ is defined as  
($V_{xx}$-$V_{yy}$)/$V_{zz}$, where $V_{zz}$ is called {\it the} electric field gradient EFG. The corresponding tensor components 
and asymmetry parameters are given in Table \ref{t_holes}. Some of the site symmetries give rise to non-diagonal components,
which are, however, very small. Therefore the principle tensor axes nearly coincide with the crystal axes and thus 
$V_{\gamma\gamma}$ are given as the corresponding diagonal elements. Recent experimental data \cite{Singer} on EFGs in 
La$_{2-x}$Sr$_{x}$CuO$_{4}$ clearly show the following observations:
There is (i) a doping-induced upward shift of the mean copper EFG  in the order of 10\%, 
(ii) a splitting of the resonance line according to different inequivalent sites, and (iii) a  
decrease in the splitting of the resonant frequencies of two different Cu sites with increasing doping. 
As one can conclude from Table  \ref{t_holes}, all these findings are consistent with our results.\cite{EFG} NMR measurements 
on Tl-based compounds \cite{Gerashenko:99} exhibit changes in the Cu and O EFGs of the order of 10\% with doping, again 
concomitant with our calculations. At the same time, 
NQR data taken on Hg1202 also revealed an upward shift with doping,\cite{Gippius:97} 
which is, however, too large in order to be understood in terms of oxygen-doping induced 
holes only. We suppose additional defects \cite{Chmaissem,Antipov:02} to be responsible for 
such a strong increase of the resonant frequencies. Our results are in a good qualitative 
agreement with the experimental observations of Ref. [\onlinecite{Bobroff97}] performed on under- and overdoped  HgBa$_2$CuO$_{4+\delta}$ crystals. In these measurements, a shift, a splitting, and a change of the in-plane $^{17}$O NMR line width were revealed with the change of the doping concentration.

\subsection{Density of states}

\begin{figure}[h]
\begin{center}
\includegraphics[angle=-90, width=1.1\columnwidth]{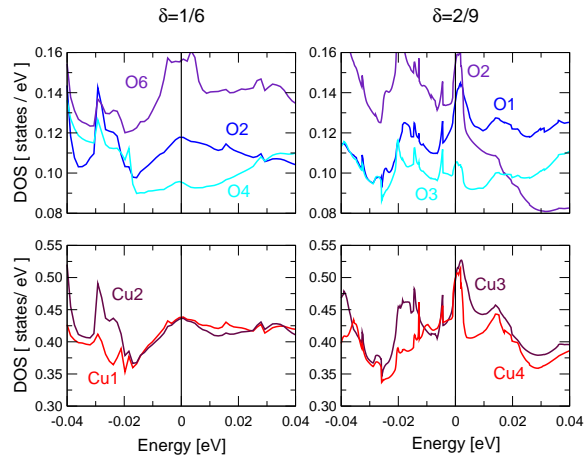}
\end{center}
\caption[]{(color online) Selected Cu($d_{x^2-y^2}$) and O($p_x$) contributions to the density of states in 
states per eV stemming from the respective atomic spheres for two different doping concentrations. The vertical lines indicate the Fermi level.}
\label{fig_dos}
\end{figure}

Having discussed the site-dependent properties, we turn our attention to the density of states (DOS), which is an integral 
characteristic of the system and should be crucial for the transition temperature.  In Fig. \ref{fig_dos}, the site-projected 
densities of states are presented for $\delta$=0.167 and 0.22, highlighting the contributions of selected copper and oxygen spheres. Like the charge carriers, the DOS's also exhibit pronounced inhomogeneities which are of similar order of magnitude. The relative differences in the DOS at the Fermi level from different oxygen sites are up to roughly 70 \% in both cases, while these are much smaller for the copper spheres. For $\delta$=0.167 the two values are nearly the same, while the differences are somewhat bigger for $\delta$=0.22, where the smallest and largest contribution are 0.468 (not shown in the figure) and 0.500, respectively. This feature is much more pronounced when moving away from the Fermi level in the range of 0.05 eV which is a typical energy scale of quasiparticles mediating the pairing. At $\delta$=0.167 all contributions show broad peaks at the same energy which is slightly below the Fermi level. For $\delta$=0.22 the peaks have become much sharper and have moved somewhat above $E_F$. Again, all copper and oxygen contributions exhibit their maxima at the same position.

Although charges and densities of states demonstrate similarities, which are inhomogeneities 
of the same order of magnitude, there is a clear difference in their behavior when we focus 
on the DOS in the vicinity of the Fermi level. This comes from the fact that the charges are
obtained by integrals of the DOS over the whole valence energy range. Yet, only a small energy range 
around $E_F$ is relevant for superconductivity.  A noticeable fact in this context is that upon doping 
the $pd \sigma$ band becomes less populated, which is important for the charges in the CuO$_2$ plane. 
Regarding the DOS at $E_F$, also the position of the saddle point (often referred to as van Hove 
singularity) with respect to the Fermi level plays a role.

\section{Conclusions}

To summarize, we have performed ab-initio calculations for the oxygen doped HgBa$_2$CuO$_{4}$ to 
consider inhomogeneity effects on the charges and densities of states. We find pronounced 
differences in the doping-induced number of holes at given copper and oxygen sites, where 
the degree of the inhomogeneity strongly depends on the doping level and the atomic species. 
Our results are consistent with the findings of Pan {\it et al.} \cite{Pan:01} giving a spatial 
scale of a few lattice parameters. Furthermore, they explain the behavior of NMR and NQR lines, 
i.e. a shift and a splitting with doping.\cite{Gerashenko:99,Singer,Gippius:97} 

Besides the atomic-scale experiments discussed above, there could be further consequences of
the non-uniform charge distribution on larger spatial scales. This distribution gives rise to 
a random potential for the carriers near the Fermi level leading to their momentum relaxation. 
On the macroscopic scale, these effects have been seen in the resistivity measurements\cite{Yamamoto01}
and electronic Raman spectroscopy.\cite{Gallais03} In a $d_{x^2-y^2}$ superconductor, such a
potential is pair-breaking and leads to a reduction of $T_c$. This decrease, however, turned out to 
be smaller than expected from the momentum relaxation rate, as it was shown in Ref.[\onlinecite{Kee01}]. 
On the nanoscale, this random potential leads to the electron inhomogeneity for the low-energy states 
probed by the scanning tunneling spectroscopy in the superconducting 
state.\cite{McElroy:05,McElroy:03,Lang02,Kinoda03}

Concerning the density of states, which is directly connected to superconductivity, we observe that the 
contributions of all plane copper and oxygen species are different in magnitude but peak at 
the same energy. Going below and above $E_F$ we find more pronounced differences which makes 
us conclude that the importance of the inhomogeneities for superconductivity strongly depends 
on the energy and character of the quasiparticle mediating the Cooper pairing. 

\bigskip 

\medskip
\noindent
{\bf Acknowledgments} \newline
This work was supported by the Austrian Science Fund, projects P13430, P14004 and M591. 
One of us (EYS) is grateful to A. Balatsky for valuable discussion and interest in this work.

\end{document}